\documentclass[runningheads]{llncs}
\providecommand{\script}{\scriptsize}
\usepackage[numbers,sort&compress]{natbib}
\usepackage{amsmath}
\usepackage{amssymb}
\usepackage{multicol}
\usepackage{comment}
\usepackage{listings}

\lstdefinestyle{prompt2}{
  basicstyle=\ttfamily\small,
  breaklines=true,
  frame=single,
  backgroundcolor=\color{gray!6},
  columns=fullflexible
}

\lstdefinestyle{prompt}{
  basicstyle=\ttfamily\script,
  breaklines=true,
  frame=single,
  backgroundcolor=\color{gray!6},
  columns=flexible,
  keepspaces=false,       
  xleftmargin=0pt,        
  framexleftmargin=0pt,   
  breakindent=0pt,        
  breakautoindent=false
}

\usepackage{orcidlink}
\newcommand\orcid[1]{\textsuperscript{\orcidlink{#1}}}

\usepackage{xcolor}
\usepackage{xspace}
\newcommand{\ld}[1]{\textcolor{purple}{LD: #1}}
\newcommand{\ey}[1]{\textcolor{blue}{EY: #1}}
\newcommand{\ww}[1]{\textcolor{red}{WW: #1}}
\renewcommand{\ld}[1]{}
\renewcommand{\ey}[1]{}
\renewcommand{\ww}[1]{}

\newcommand{\autoargue}[0]{{\sc Auto\-Argue}}

\newcommand{\crucible}[0]{{\sc Crucible}}
\newcommand{\cruciblefiltered}[0]{\crucible{}{\sc{-Verified}}}
\newcommand{\cruciblebase}[0]{\crucible{}{\sc{-Base}}}
\newcommand{\ginger}[0]{{\sc Ginger}}
\newcommand{\gingerllama}[0]{{\sc Ginger-Llama}}
\newcommand{\gptresearcher}[0]{{\sc GptResearcher}}
\newcommand{\bulletpoints}[0]{{\sc BulletPoints}}
\newcommand{\LSR}[0]{MILCO}

\newcommand{\llama}[0]{LLaMA}
\newcommand{\llamabig}[0]{\texttt{\llama{}-\-3.3-\-70B-\-Instruct}}

\newcommand{\down}[0]{$^\triangledown$}
\newcommand{\up}[0]{$^\blacktriangle$}

\usepackage{booktabs}
\usepackage{makecell}
\usepackage{bbding}

\newcommand{\tightgraph}[1]{\paragraph{#1}}

\begin{document}

\title{Incorporating Q\&A Nuggets into Retrieval-Augmented Generation}
\titlerunning{Incorporating Q\&A Nuggets into Retrieval-Augmented Generation}
%
\author{Laura Dietz\inst{1}\orcid{0000-0003-1624-3907} \Envelope \and
 Bryan Li\inst{2}\orcid{0000-0002-5779-1662} \and
 Gabrielle Liu\inst{3}\orcid{0000-0002-0603-1655} \and
 Jia-Huei Ju\inst{4}\orcid{0000-0003-2247-3370} \and
 Eugene Yang\inst{5}\orcid{0000-0002-0051-1535} \and
 Dawn Lawrie\inst{5}\orcid{0000-0001-7347-7086} \and
 William Walden\inst{5}\orcid{0000-0001-9931-2861} \and
 James Mayfield\inst{5}\orcid{0000-0003-3866-3013}}
\authorrunning{Dietz et al.}

\institute{
University of New Hampshire, Durham, New Hampshire, USA \email{dietz@cs.unh.edu} 
\and
University of Pennsylvania, Philadelphia, Pennsylvania, USA \email{bryanli@seas.upenn.edu} 
\and
Yale University, New Haven, Connecticut, USA \email{kaili.liu@yale.edu}
\and
University of Amsterdam, Amsterdam, Netherlands \email{j.ju@uva.nl} 
\and
Human Language Technology Center of Excellence, Johns Hopkins University, Baltimore, Maryland, USA \\
\email{\{eugene.yang,lawrie,wwalden1,mayfield\}@jhu.edu}
}

\maketitle           

\begin{abstract}
RAGE systems integrate ideas from automatic evaluation (E) into Retrieval-augmented Generation (RAG).
As one such example, we present \crucible{}, a Nugget-Augmented Generation System that preserves explicit citation provenance by constructing a bank of Q\&A nuggets from retrieved documents and uses them to guide extraction, selection, and report generation. 
Reasoning on nuggets avoids repeated information through clear and interpretable Q\&A semantics---instead of opaque cluster abstractions---while maintaining citation provenance throughout the entire generation process.
Evaluated on the TREC NeuCLIR 2024 collection, our \crucible{} system substantially outperforms \textsc{Ginger}, a recent nugget-based RAG system, in nugget recall, density, and citation grounding.
\footnote{Appendix at \url{https://github.com/hltcoe/ecir26-crucible-system-appendix/}}
\end{abstract}


\section{Introduction}
\ld{STOP!! disable todo notes in preamble before submitting!}
Retrieval-Augmented Generation (RAG) has become the dominant
framework for grounding LLM outputs in evidence
\cite{lewis2020retrieval,izacard2021leveraging}.
At the same time, nugget-based evaluation methods have emerged as the standard for measuring relevance of long-form responses \cite{mayfield2024evaluation,farzi2024pencils}.
A \emph{nugget}, i.e., short Q\&A pair, fact, or claim, is a fine-grained reusable unit for assessing whether key information is covered. 
By encoding the information that must appear in a system answer, nuggets enable evaluation metrics such as ``nugget recall'' that directly quantify the amount of useful information given by the information system.
We argue that nuggets are valuable not only for evaluation but also for guiding retrieval and generation, especially given that LLMs can produce high‑quality nuggets automatically \cite{dietz2024workbench,farzi2024exampp}.

\tightgraph{Contributions.}
To support this argument, we present \crucible{}, a nugget-centric RAG system that automatically constructs its own nugget bank and uses it as a control signal throughout the pipeline. For example, nuggets form the basis for controlling redundancy without content-based clustering, while naturally preserving citation provenance. 
On the TREC NeuCLIR 2024 test collection, we provide a direct comparison to a recent nugget-oriented RAG system \ginger{} \cite{lajewska2025ginger}. We find  that \crucible{} decisively outperforms \ginger{} across several evaluation metrics, from nugget recall to citation support.


\section{Related Work}

\tightgraph{Summarize one‑document‑at‑a‑time.}
Early RAG systems either produce a summary per retrieved passage \cite{lewis2020retrieval} or encode each passage separately \cite{izacard2021leveraging}.  This retains exact citation provenance, but the model must attend to many independent inputs, which often harms coherence and fluency.  

\tightgraph{Joint representation.}
Later work uses a single joint representation to represent the retrieved document set. \textssc{RealM} \cite{guu2020realm} and \textsc{EMDR2} \cite{sachan2021emdr2} learn a unified encoding; \textsc{xRAG} \cite{cheng2024xrag} pushes this to the extreme by compressing all content into a single token representation.  These approaches improve synthesis and fluency, yet the compact latent representation discards explicit links to source documents, making citation grounding difficult and the pipeline opaque to developers.  
A recommended remedy is to first hallucinates an answer and then retrieve supporting evidence, as in \textsc{HyDE} \cite{gao2023precise}, which raises lingering concerns about trustworthiness.

\tightgraph{Clustering‑based RAG.}
%
\ginger{} \cite{lajewska2025ginger} is a nugget-based RAG system that defines nuggets as verbatim text spans copied directly from retrieved passages. These spans are first clustered into topical facets using BERTopic, after which the clusters are reranked to identify the most relevant facets. Summaries of the top clusters form the system’s output, which is then rewritten for fluency. 

Although the pipeline has access to evidence for generation, the clustering summarization step abstracts away from the original document extractions, impacting the faithful citation grounding.


\tightgraph{Agentic RAG.}
Agentic frameworks decompose the pipeline into subtasks and let a planner invoke them as needed \cite{yang2024rag,asai2023selfrag}.  \textsc{WebGPT} \cite{nakano2021webgpt} embeds search instructions in the prompt, letting the LLM choose passages to cite.  A prominent agentic system is \gptresearcher{} \cite{duh2025hltcoe-liverag,Elovic_gpt-researcher_2023}, which orchestrates query generation, retrieval, extraction, and report writing, with optional verification and trace‑back.

\section{Nugget-first RAG Approach: \crucible{}}
\label{sec:approach}

\begin{figure}[t]
    \centering
    \includegraphics[width=1\linewidth]{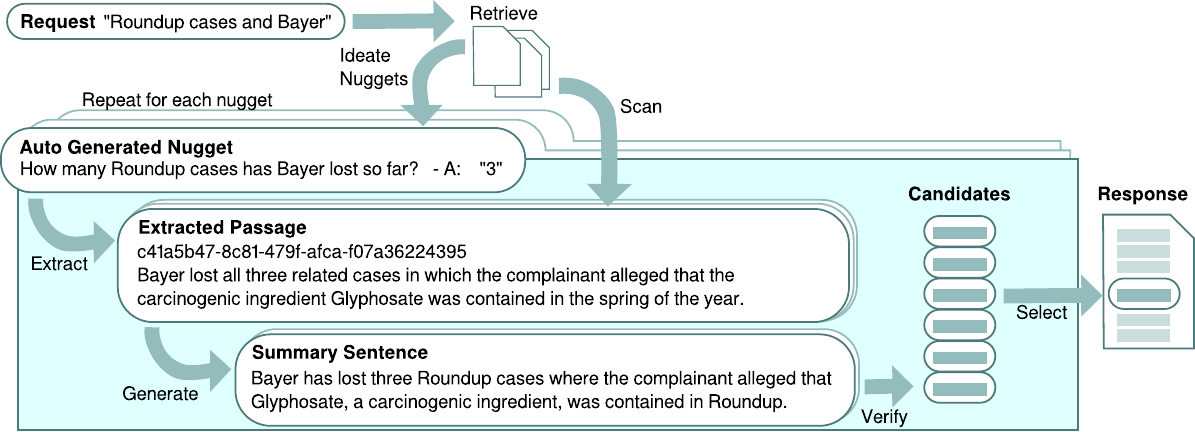}
    \caption{ 
   For each generated nugget, \crucible{} extracts candidate sentences and adds the best $k$ sentences  to the final response. No content clustering is needed.}
    \label{fig:crucible}
\end{figure}



Figure~\ref{fig:crucible} illustrates the workflow of \crucible{}, which
builds its pipeline around structured Q\&A nuggets that guide
retrieval, extraction, and assembly.  This contrasts with \ginger{}, which
operates on verbatim spans clustered into facets.

\tightgraph{Nugget ideation.} We begin by retrieving an initial pool of documents; in our study we use the PLAID‑X dense retriever \cite{yang2024translate}.  
For each document we generate a short query-focused summary with an LLM, then prompt the
same LLM (using \llamabig{}) to produce Q\&A nuggets conditioned on the
summary and the user request.

To reduce redundancy among nuggets, we first detect paraphrases with an LLM prompt, then successively merge most confident paraphrases until the desired number of canonical nuggets  is obtained.

\tightgraph{Nugget ranking.} 
The resulting nugget bank is reranked with a Support‑Vector
Classifier (SVC) trained on 19 quality features, which utilize LLM judge prompts to measure how well each nugget addresses the information need (task statement, background, role, communication style, and scope), degree of vitality \cite{10.1145/3726302.3730090}, and several features from researchy questions \cite{rosset2024researchy}. This is combined with basic readability metrics to quantify reading level \cite{textstat} and sentence complexity \cite{liu2023makes}.
Using the number of paraphrases as an indicator for nugget quality, we fuse this feature-based ranking with a popularity-based ranking. Obtaining the top 20 nuggets for each request will form the nugget bank that drives the remainder of the response generation.

\begin{figure}
\centering
\begin{minipage}{1\linewidth}
\lstset{style=prompt}
\begin{scriptsize}
\begin{lstlisting}
Given the following question and answer, please find the sections in the provided source document that support and validate the answer to the question. 
If the source document does not support the given answer, then set confidence to 0.0 and all fields to None.
Provide the section from the document that supports the answer, with complete sentences directly from the document. Please include context around each supporting segment, making sure that there is enough context to support why the answer is a correct response to the question. Your response should include just the extracted text segment, and nothing else.
Then condense the extracted text into one concise sentence that clearly demonstrates how the question is answered, without referring to the source document.

- Emit **each field exactly once**, in the order shown.
- Do NOT repeat headers. Do NOT include any field more than once.
- If a field is unknown, omit it entirely - do NOT write `None` or `null`.
\end{lstlisting}
\end{scriptsize}
\end{minipage}

\begin{scriptsize}

\vspace{0.8em}
\begin{tabular}{@{}lp{0.68\linewidth}@{}}
\toprule
\textbf{Field name} & \textbf{Description}\\ \midrule
In: \texttt{nugget\_text} & Input question\\
In: \texttt{answer} & Valid answer or list of answers\\
In: \texttt{source\_document} & Input document chunk\\
In: \texttt{title} & Title query\\
In: \texttt{background} & User background\\
In: \texttt{problem\_statement} & User's problem statement\\
Out: \texttt{extracted\_text\_segment} & Passage returned by the model\\
Out: \texttt{summary} & Single concise sentence drawn from the passage\\
Out: \texttt{reasoning} & Internal chain of thought (ignored downstream)\\
Out: \texttt{confidence} & Model reported confidence value (float)\\
\bottomrule
\end{tabular}
\end{scriptsize}
\caption{Prompt for scanning, extraction, and sentence candidate generation.
\label{prompt:Answer}}
\end{figure}


\tightgraph{Retrieval.}
Our base system uses documents of which nuggets were extracted. Additionally, the
pipeline is evaluated with the top 100 of several retrieval models with their default configurations: PLAID‑X \cite{yang2024translate} a dense multilingual retriever, using top 25 per language, dense retrieval with Qwen3 \cite{zhang2025qwen3}, and Milco \cite{nguyen2025milco}.

\tightgraph{Scanning and generation.} Using the nugget bank, we scan retrieved documents\footnote{Segmented into 1000 character chunks, split at sentence boundary.} for passages that directly answer each nugget.  Using the prompt in Fig.\ \ref{prompt:Answer}\footnote{Unlike a previous manuscript version, the results here correctly use the prompt.} we (1) locate a supporting
passage, (2) extract a concise self‑contained sentence, and (3) record
the LLM's token‑likelihood as the extraction confidence.  

\tightgraph{Verification (optional).} Extractions may be double‑checked for nugget coverage
and citation support.  This step invokes prompts used by the \autoargue{} system \cite{walden2025autoargue} albeit with the automatically-generated nuggets identified after the Nugget Ranking step; the LLM is
prompted with a binary ``YES/NO'' question asking whether a candidate sentence
is truly supported by the cited span or whether the nugget is truly covered by the candidate sentence and extracted passage. 
As this step risks leaking evaluator knowledge into the system, 
we also evaluate the system with this step skipped as ``\cruciblebase{}''.

\textbf{Sentence selection.} We rank the remaining candidates for each nugget by the extraction confidence and choose the top $k$
sentences ($k=1$ herein; results for larger $k$ are in the
online appendix). This ensures that each sentence is tied to exactly one citation.

\textbf{Assembly.} Selected sentences are concatenated into a report,
ordered by nugget quality ranking. Sentences with same stopped/stemmed text are omitted.  Because every sentence is self‑contained and atomic, the
order of sentences does not affect the readability.
Every sentence cites exactly one document.

\section{Evaluation Setup}
\paragraph{Dataset.}
We evaluate on the TREC NeuCLIR 2024 Report Generation Pilot~\cite{lawrie2024overview}, which requires drafting topic‑focused reports supported by citations drawn from a multilingual corpus, where all language subsets are machine‑translated to English. The topics are broad, e.g., ``Roundup cases and Bayer'', yet each report must be tailored to the specific problem statement and user background.
%
 The collection's evaluation tool, \autoargue{} \cite{walden2025autoargue}, uses held‑out, manually curated gold‑nugget banks.

\paragraph{Compared systems.}
Using \llamabig{} and PlaidX retriever, we compare two variants of \crucible{} with state-of-the-art systems:
\begin{description}
    \item[\cruciblebase{} (ours),] as described in Section \ref{sec:approach} without the verification step.
    \item[\cruciblefiltered{} (ours),] full pipeline with check for coverage/support.
    \item[\ginger{} \cite{lajewska2025ginger},] a nugget-informed RAG system that generates responses based on clusters built on the extracted nuggets from the retrieved documents. While the released implementation does not produce citations in its response,
    we assign citations to the document that contains the nuggets closest to the cluster centroid.
    This is an alternative system that is based on ideas similar to those in \crucible{}.
    We report the best variant, where initial set of documents is retrieved by Qwen3-Embedding bi-encoder~\cite{zhang2025qwen3}, underlying LLM is \texttt{GPT-4o}. 
    \item[\gingerllama{}:] Using the GINGER implementation with the same \llama{} LLM
    to allow a fair comparison to \crucible{}.

%
    \item[\gptresearcher{}~\cite{duh2025hltcoe-liverag,Elovic_gpt-researcher_2023}] a simple agentic system with subqueries, retrieval, and writing. Uses \llama{} as LLM (despite the name).
%
    \item[{\sc BulletPoints}~\cite{yang2025hltcoetrec},] an extractive pipeline (best in TREC NeuCLIR 2024, submitted as \texttt{hltcoe‑eugene}) which uses \llama{} as LLM.
\end{description}

\paragraph{Evaluation metrics.}
We report metrics from the \autoargue{} framework \cite{walden2025autoargue}.

\begin{description}
    \item[Nugget Recall] $=\frac{\text{covered nuggets}}{\text{gold nuggets}}$ measures the recall of gold nuggets.
    \item[Nugget Density] $=\frac{\text{covered nuggets}}{\text{sentences}}$ balance of coverage vs.\ concise summaries. 
    \item[Sentence Novelty] $=\frac{\text{sentences mention a new nugget}}{\text{all sentences}}$, fraction of sentences that introduce new relevant information, according to gold nuggets.  
    \item[Relevant Sentences] $=\frac{\text{sentence with nuggets}}{\text{all sentences}}$ measures how many sentences mention relevant nuggets.
    \item[Citation Support] $=\frac{\text{supported citations}}{\text{citations}}$, 
    measures how many citations actually support their summary sentence.
\end{description}

\section{Results}

\tightgraph{Overall performance.}
Table~\ref{tab:results} reports results averaged across NeuCLIR topics.
\crucible{} consistently outperforms \ginger{} and \ginger{}-LLaMA on all
nugget-oriented metrics. Relative gains are especially pronounced ranging between ($+56$\%) for \emph{nugget density} and  ($+267$\%) for \emph{relevant sentences}.
We attribute \crucible{}'s gains to its nugget-first design: explicit
ideation yields systematic coverage; per-nugget extraction enforces
grounding; and fingerprint duplicate checks preserve density. In contrast,
\ginger{}’s cluster-based summarization risks citation traceability, while not providing the required information. We remark that GINGER was designed for the TREC RAG 24 \cite{upadhyay2024llm} task and the Autonuggetizer \cite{pradeep2024autonuggetizer} for evaluation.

\begin{table}[t]%
\centering%
\caption{Comparison of RAG systems (each with PlaidX retriever) on TREC NeuCLIR 2024. Best results in bold, paired t-test \up{}/\down{} with reference \textsc{Base}.}%
\label{tab:results}%
\begin{footnotesize}%
\begin{tabular}{lrrrrr}%
\toprule%
System&\makecell{Nugget\\Recall}&\makecell{Nugget\\Density}&\makecell{Sentence\\Novelty}&\makecell{Relevant\\Sent.}&\makecell{Citation\\Support}\\%
\midrule%
\cruciblebase{}&\textbf{0.584}&0.392&0.247&0.805&0.595\\%
\cruciblefiltered{}&0.562&\up{}\textbf{0.495}&\up{}\textbf{0.330}&\up{}\textbf{0.826}&\up{}\textbf{0.932}\\%
\midrule%
\gptresearcher{}&\down{}0.177&\down{}0.131&\down{}0.083&\down{}0.265&\down{}0.571\\%
\ginger{}&\down{}0.230&\down{}0.251&\down{}0.149&\down{}0.219&\down{}0.450\\%
\ginger{}-LLaMA&\down{}0.167&\down{}0.108&\down{}0.086&\down{}0.127&\down{}0.438\\%
\textsc{BulletPoints}&\down{}0.508&\down{}0.340&0.243&\down{}0.468&\up{}0.835\\%
\bottomrule%
\end{tabular}%
\end{footnotesize}%
\end{table}

\begin{table}[t]%
\centering%
\caption{Downstream effects of the document retrieval method and optional verification; using the same nugget bank across variations; reference \crucible{}. Paired t-test with respect to the "From nuggets" in each row set.}%
\label{tab:results-retrieval}%
\begin{footnotesize}%
\begin{tabular}{llrrrrr}%
\toprule%
&Retrieval&\makecell{Nugget\\Recall}&\makecell{Nugget\\Density}&\makecell{Sentence\\Novelty}&\makecell{Relevant\\Sent.}&\makecell{Citation\\Support}\\%
\midrule%
\cruciblebase{}&From nuggets&0.599&0.400&0.261&0.768&0.666\\%
&\LSR{}&0.591&0.393&0.259&0.764&0.641\\%
&Qwen3&\textbf{0.616}&\textbf{0.417}&\textbf{0.275}&0.778&\textbf{0.673}\\%
&Plaid-X&0.584&0.392&0.247&\up{}\textbf{0.805}&\down{}0.595\\%
\midrule%
\cruciblefiltered{}&From nuggets&\textbf{0.607}&0.446&0.287&0.793&0.830\\%
&\LSR{}&\down{}0.559&0.452&0.305&0.810&\up{}0.959\\%
&Qwen3&0.600&\up{}0.493&\up{}0.326&0.810&\up{}\textbf{0.960}\\%
&Plaid-X&\down{}0.562&\up{}\textbf{0.495}&\up{}\textbf{0.330}&\up{}\textbf{0.826}&\up{}0.932\\%
\bottomrule%
\end{tabular}%
\end{footnotesize}%
\end{table}

 \gptresearcher{} is similarly strongly outperformed. The closest contender is \bulletpoints{} for which a significant improvement is obtained for \emph{nugget recall}, \emph{nugget density}, and \emph{relevant sentences}. Merely in terms of citation support, \bulletpoints{} ourperforms our \cruciblebase{} model.
 
\tightgraph{Ranking model.}
\crucible{} obtains its great performance under all explored retrieval models in Table \ref{tab:results-retrieval}. We note similar results on \ginger{}. Only on \cruciblefiltered{}, PlaidX performs better.

\tightgraph{Verification step.}
Using \autoargue{}'s prompts to verify the coverage of system nuggets can obtain improvements when evaluated against gold nuggets, especially for the nugget density  (+26\%) and sentence novelty (+33\%) metrics.

In contrast, the citation support increases by (+64\%) to a near-perfect score of 0.932. After manual inspection, we believe that it is not that original citations were incorrect, but during verification we prefer sentences where the support is clearly articulated in the extracted document passage, 
which is ultimately helpful for the user to trust the system.

\tightgraph{More experiments.} 
\crucible{} runs were submitted to several TREC 2025 tracks: DRAGUN, RAG, RAGTIME \cite{dietz2025hltcoe}. While many relevance judgments are not available at the time of writing, we have some preliminary corroborating evidence.

\tightgraph{Costs.}
This system was designed to study possible quality improvements, not to obtain a fast system ready for deployment.
%
The runtime cost is dominated by LLM calls. One of the most expensive steps is the paraphrasing detection of the nugget ideation with $\Theta(D^2)$ prompts, which could be scaled with finger printing and SimHash approaches. Other expensive steps are nugget scanning with $\Theta(N\cdot D)$ LLM calls and verification with $\Theta(S)\approx \Theta(N \cdot D)$ LLM prompts, these could be made more efficient by tracking the quality of extractions, and stopping when $k$ sentences of sufficiently high quality have been found.

\tightgraph{Limitations.} \crucible{} is developed working closely with organizers of TREC NeuCLIR and developers of the \autoargue{} evaluation system. As shown in \citet{dietz2026insider}, this knowledge can lead to a distortion of empirical evaluation results. Since, the verification phase, represents a circularity with \autoargue{}, we offer results with and without this phase. It is quite possible that near-perfect citation-support scored are due to the circularity between our verification step and the \autoargue{} evaluation system. 
We further study this potential vulnerability in the upcoming TREC 2026 Auto-Judge track.

\section{Conclusion}
We demonstrate one example of how RAGE systems, which incorporate ideas from automatic evaluation (E) into RAG systems, can lead to significantly better performing systems. 
To substantiate this claim we introduce \crucible{}, a nugget‑first pipeline that uses a bank of Q\&A nuggets to guide retrieval, extraction, and final assembly. Optionally, \crucible{} uses LLM-judge prompts as a verification steps.  On the NeuCLIR benchmark \crucible{} decisively outperforms \ginger{}, demonstrating the practical benefits controlling the generation.

The key insight behind \crucible{} is that citation provenance can be lost whenever a system condenses raw text into latent or clustered summaries.  Most contemporary RAG approaches either (1) struggle to avoid redundancy (e.g. FiD \cite{izacard2021leveraging}) or (2) sever the  explicit link between a generated sentence and its source document.  \crucible{} avoids both pitfalls by first constructing a set of canonical Q\&A nuggets (any required clustering occurs at this stage) and only then extracting and summarizing sentences that address each nugget. The generated report will contain the best sentence for each nugget, each sentence is supported with exactly one citation. Obtaining near-perfect citation support score of 0.9 (\cruciblebase{}) and 0.96 (\cruciblefiltered{}) supports that either this solves the citation-support problem, or that we need more precise evaluation methods for citation support.

We hope that future work will explore a greater variety of RAGE systems across the gamut of RAG and LLM-as-a-Judge paradigms.  

\paragraph{Disclosure of Interests.} Authors have no competing interests.

\bibliographystyle{splncs04nat}
\bibliography{bibio}

\end{document}